\documentclass[aps,twocolumn,groupedaddress]{revtex4}
\usepackage{amsfonts}
\usepackage{amsmath}
\usepackage{amssymb}
\usepackage[dvipdfmx]{graphicx}
\usepackage{braket}
\usepackage{amsmath}
\usepackage{amsfonts}
\usepackage{amssymb}
\usepackage{color}%

\allowdisplaybreaks[3]

\providecommand{\U}[1]{\protect\rule{.1in}{.1in}}
\providecommand{\U}[1]{\protect\rule{.1in}{.1in}}
\begin{document}
\title[Required number of states increases only moderately with the problem size for antisymmetrized geminal powers]
{\textbf{Required number of states increases only moderately with the problem size for antisymmetrized geminal powers}}
\author{Wataru Uemura and Takahito Nakajima}

\affiliation{RIKEN Center for Computational Science, Kobe 650-0047, Japan}
\pacs{PACS number}

\begin{abstract}
We propose an algorithm to obtain the 
ground-state energy of a many-electron system
using the variational wave function
of a linear combination of antisymmetrized geminal powers.
We optimized this algorithm to
obtain the energy and the other parameters of a many-electron system.
Also we clarified the bottleneck of the total
calculation in the tensor contraction and successfully reduced
the computational time. 
As a result, we can use 
an extended number of geminal states to
obtain the ground state of the water molecule 
and Hubbard models.
The result for the water molecule with the Dunning double-zeta basis is of the sub-milihartree order above the energy of exact diagonalization.
Further, we observe that the result for the one-dimensional Hubbard model with 14 sites shows good tendency to capture the right ground state and that for the two-dimensional Hubbard 
model still lacks some part of the energy
reflecting the large size of the Hilbert space.
We conclude that the required number of 
terms for geminal states
for sufficiently accurate energy 
is only moderately affected by the 
problem size.
We further show other technical details for 
the numerical algorithms of geminal states in 
the variation process.
It is expected that with the 
use of more extended computing 
resources and larger sizes of electronic systems, 
our algorithm can provide improved results.

\end{abstract}

\maketitle

\section{Introduction}

The many-electron problem 
in quantum chemistry or
quantum lattice models is still
considered as an important subject of  
modern science.
With the growth of large-scale 
computer facilities, 
it has become possible to not only 
solve these problems with 
the exact diagonalization but also
construct 
numerical algorithms that could handle
the complex behaviors of 
electronic correlation.\par
In post-Hartree-Fock theories,
perhaps the most simple and 
convenient method is configuration interaction.\cite{shavitt, saxe}
Configuration interaction
is built on the basis of the Hartree-Fock 
theory, and a small or large number of  configurations are chosen for 
each approximation level.
Obtaining
energy results of high precision
is limited by the large 
scaling problem of the algorithm.\par
The configuration interaction
or coupled cluster\cite{purvis} is regarded
as a single reference that is based 
on the results of one Hartree-Fock
configuration.
There is already an established 
framework for multi-reference 
configuration interaction (MRCI).\cite{knowles}
MRCI uses several configurations
for the building blocks and 
applies excitation or other operators
to these configurations.
Thus, the MRCI can
obtain results that are very close to 
the exact result; however, 
the calculations become 
prohibitively expensive 
for large molecules.
The multi-reference coupled cluster (MRCC) is also used in such applications, but it is yet early to provide any systematic 
applications.\cite{Adamowicz, evangelista}\par
In the area of macroscopic 
materials, the most 
successful method for the 
many-electron problem 
is density functional theory (DFT).\cite{hohenberg, kohn, levy}
The formalism of DFT is based on the 
assumption that there is a one-to-one correspondence between the total wave function and the single-electron density.
The Kohn-Sham equation is built 
on this theory and treats the system
on the basis of the Hartree-Fock 
theory.
If we can replace the Hartree-Fock 
strategy in DFT with a more sophisticated one with 
multi-reference wavefunction
ansatz, then we can possibly obtain 
a large part of the correlation energy
of the target system.\par
When solving the many-electron problem with
multi-reference theory,
quantum Monte Carlo (QMC) is a highly 
reliable tool.\cite{foulkes, bajdich2006, bajdich2008, eric, eric2, sorella, sorella2, sorella3, booth, haoshi, imada}
It is common to combine the 
Slater determinant with the 
Jastrow factor to obtain
the solution of a target system.
In some cases, the wavefunction
is extended to use the Pfaffian
or antisymmetrized geminal power (AGP) states to obtain an accurate
ground state.
There are known problems in 
QMC; it suffers from
the well-known negative-sign
problem with the usage of probability.\par
The formalism of configuration 
interaction or MRCI is 
not dependent on the probability and is instead based on 
a discrete set of configurations.
If we construct the Slater
determinants from 
continuous matrices, the 
degree of freedom for the 
variation is extended considerably.
On this subject, there are foregoing 
research works mainly on 
Hubbard models.\cite{fukutome, tomita1, tomita2, tomita3}
The formalism of their calculations
is based on the evaluation
of the matrix element of the 
overlap of
the multi-Slater determinant 
with matrices representing 
the orbitals on each site.
Because of the nature of the 
Hubbard model that has a simple form of the 
correlation term,
the calculation for  
Hubbard models with 
non-orthogonal determinants 
could be achieved with low computational
scaling.
With this formalism, 
the ground states of one-dimensional
and two-dimensional Hubbard models
are captured adequately and with good precision
of the energy.
Recently, new results have been reported  that are aided by large computing 
facilities.\cite{hoyos1, hoyos2, hoyos3}
These report highly accurate results
with the consideration of 
the spin and spatial symmetry
for the Hubbard model as well as
qualitatively good results for the selected molecular
systems.\par
The electronic states of the 
Slater determinants are 
classified as non-interacting, and 
they are described by rectangular
matrices and their inverse matrices.
There are suggestions in
the context of reduced 
density matrices for the fermionic 
system to use the 
antisymmetrized geminal power states
instead of the Slater determinants.\cite{coleman1, coleman2, Ortiz1981, weiner, giorgini, Straroverov2002, Scuseria2011}
We have recently shown 
the formula for the density matrices for 
geminal states and variational 
calculations with the superpositions
of geminal states.\cite{uemura2015, uemura201901}
The result were, in some cases, highly accurate
when compared with 
the exact diagonalization.
An algorithm was included to avoid 
the divergent-like behavior 
of the AGP energy.
We have extended 
the number of states for the AGP
and tested the same molecular system
and the Hubbard model with increased number of electrons.
The results are especially accurate
in the case of a moderate number of geminal states.
From these observations,
we concluded that the 
exact ground-state wavefunction
could be described with
a moderate number of states 
for multi-reference AGP ansatz.
In the next section, we provide 
some algorithms for the 
energy variation of the AGP states.

\section{Formalism}
First, we introduce the derivation of the expression for 
energy with two AGP states.
This result was first indicated in
ref.\cite{uemura2015}.
When constructing the AGP state
from matrix parameters, we use the permutation tensor
\begin{equation}
\epsilon _{i_1\cdots i_N} = \hat{A}(\epsilon _{i_1i_2} \epsilon _{i_3i_4} \cdots \epsilon_{i_{N-1}i_N}). \label{permutation geminal}
\end{equation}
The matrix $\epsilon $ is defined 
in ref.\cite{uemura2015} as a skew-symmetric matrix.
From the permutation tensor, 
we can construct the AGP 
wavefunction as
\begin{equation}
A _{i_1\cdots i_N}=\sum _{k_1\cdots k_N} 
\epsilon _{k_1\cdots k_N} 
\alpha _{k_1i_1} \cdots \alpha _{k_Ni_N}.
\end{equation}
Here, $\alpha $ is a general matrix.
We also define the superposition
of the AGP states as
\begin{equation}
A _{i_1\cdots i_N}=\sum _{k=1}^{K}c_k\sum _{k_1\cdots k_N} 
\epsilon _{k_1\cdots k_N} 
\alpha ^k_{k_1i_1} \cdots \alpha ^k_{k_Ni_N},
\end{equation}
where $c_k$ is a variational parameter.
This representation with $\alpha $
is equivalent to the AGP states 
with geminal $\gamma $ by
the relationship
\begin{equation}
\gamma _{i_1i_2} = \sum _{k_1k_2}\epsilon _{k_1k_2}\alpha _{k_1i_1}\alpha _{k_2i_2}. \label{gamma}
\end{equation}
From matrix $\alpha $, we can
construct the overlap of the 
AGP state as
\begin{equation}
n=\sum _{i_1\cdots i_Nj_1\cdots j_N}
\epsilon _{i_1\cdots i_N} \epsilon _{j_1\cdots j_N}
a _{i_1j_1} \cdots a_{i_Nj_N} \label{norm},
\end{equation}
where
\begin{equation}
a=\alpha ^t\alpha.
\end{equation}
Expression $^tx$ shows the 
matrix transpose of matrix $x$.
When we substitute the permutation tensor $\epsilon $ of eq.(\ref{permutation geminal})
in eq.(\ref{norm}), we can obtain
the explicit formula for the AGP norm as
\begin{equation}
n=3(\mathrm{tr}\,A)^2-6(\mathrm{tr}\,A^2)
\end{equation}
for $N=4$, and
\begin{equation}
n=15(\mathrm{tr}\,A)^3
-90(\mathrm{tr}\,A)(\mathrm{tr}\,A^2)
+120(\mathrm{tr}\,A^3)
\end{equation}
for $N=6$.
Here, matrix $A$ is defined as
\begin{equation}
A=a\epsilon ^ta ^t\epsilon .\label{defA}
\end{equation}
These equations are similar to the ones that are obtained in the context of nuclear physics.\cite{mizusaki}
We can obtain the general result as
\begin{eqnarray}
n&=&N!\,\mathrm{exp}(\frac{1}{2}\mathrm{tr}\,\mathrm{log}(1+At))|_{t^{N/2}} \label{exp} \\
&\equiv &N!\,\mathrm{pf}(1+At)|_{t^{N/2}},
\end{eqnarray}
which is equivalent to the norm 
expression given in ref.\cite{uemura2015}.
The subscript $t^{N/2}$ represents the $N/2$-order coefficient of $t$.\par
Next, we introduce 
the method to obtain the second-order reduced density matrix for the AGP state.
We can use the second derivative of the 
norm as
\begin{eqnarray}
&&\frac{\partial }{\partial a_{k_1l_1}}\frac{\partial }{\partial a_{k_2l_2}}
\sum _{i_1\cdots i_Nj_1\cdots j_N}
a_{i_1j_1}\cdots a_{i_Nj_N}\epsilon _{i_1\cdots i_N}\epsilon _{j_1\cdots j_N}\nonumber \\
&=&N(N-1)
\sum _{i_1\cdots i_Nj_1\cdots j_N}
\delta _{i_1k_1}\delta _{j_1l_1}\delta _{i_2k_2}\delta _{j_2l_2}
a_{i_3j_3}\cdots a_{i_Nj_N}\nonumber \\
& &\cdot \,\epsilon _{i_1\cdots i_N}\epsilon _{j_1\cdots j_N} \nonumber \\
&=&N(N-1)
\sum _{i_3\cdots i_Nj_3\cdots j_N}
a_{i_3j_3}\cdots a_{i_Nj_N}\nonumber \\
& &\cdot \,\epsilon _{k_1k_2i_3\cdots i_N}\epsilon _{l_1l_2j_3\cdots j_N},
\end{eqnarray}

\begin{equation}
\Gamma _{i_1j_1i_2j_2}=
\sum _{k_1l_1k_2l_2} \alpha _{k_1i_1}\alpha _{l_1j_1}\alpha _{k_2i_2}\alpha _{l_2j_2}
\frac{\partial }{\partial a_{k_1l_1}}\frac{\partial }{\partial a_{k_2l_2}}n.
\end{equation}
The tensor $\Gamma $ would provide
the reduced density matrix that 
we are seeking.
We can use the relationship between matrix $\alpha $ and $\gamma $ to obtain the 
reduced density matrix formula with the 
expression $\gamma $.
These results appear to be similar to 
the expressions provided in ref.\cite{uemura2015}.\par
We can obtain the explicit formula of the first-order derivative of the total energy by simple differentiation with
respect to parameter $\gamma $.
The result is given as follows:
\begin{eqnarray}
& &\frac{\partial }{\partial \gamma ^x_{ij'}}\gamma ^x_{j'j}E^{xy}\nonumber \\
&=&\sum _{k_1l_1k_2l_2}H_{k_1l_1k_2l_2}\cdot \nonumber \\
& &(\frac{1}{2}(Q_rM\times Q_rM\times Q_rM\, q3\, Q_l\times Q_l\times Q_l)_{jl_1l_2ik_2k_1}\nonumber \\
&-&(k_1\leftrightarrow k_2)\nonumber \\
&+&\frac{1}{2}(Q_rM\times Q_r\times \gamma ^xQ_r\, r3\, Q_l\times Q_l\gamma ^y\times Q_l)_{jl_1k_2il_2k_1}\nonumber \\
&-&((Q_r\times \gamma ^xQ_r\times Q_rM\, r3\, Q_l\gamma ^y\times Q_l\times Q_l)_{l_1jl_2ik_2k_1}\nonumber \\
&-&(Q_rM \times Q_r \times Q_rM\, r3\,Q_l\times Q_l\times Q_l)_{l_1il_2jk_2k_1})\nonumber \\
&-&(k_1\leftrightarrow k_2)\nonumber \\
&+&\frac{1}{2}(Q_r\times Q_rM\times \gamma ^xQ_r\, r3\, Q_l\gamma ^y\times Q_l\times Q_l)_{l_1jk_2il_2k_1}\nonumber \\
&+&\frac{1}{2}(Q_r\times Q_r\times \gamma ^xQ_r\, s3\, Q_l\gamma ^y\times Q_l\times Q_l)_{l_1ijl_2k_2k_1}).\label{first}
\end{eqnarray}
Here, we are using the eigendecomposition 
\begin{equation}
B=Q_r\, M\, Q_l,
\end{equation}
where
\begin{equation}
B=-\gamma ^y\gamma ^x.
\end{equation}
Tensors $q3$, $r3$, and $s3$ are rank-$3$
tensors defined as follows:
\begin{eqnarray}
q3_{i_1i_2i_3} &=&
((1+Mt)^{-1}_{i_1i_1}(1+Mt)^{-1}_{i_2i_2}(1+Mt)^{-1}_{i_3i_3}\nonumber \\
&\cdot &\mathrm{pf}(1+Bt))|_{t^{N/2-3}},
\end{eqnarray}
\begin{eqnarray}
r3_{i_1i_2i_3} &=& ((1+Mt)^{-1}_{i_1i_1}(1+Mt)^{-1}_{i_2i_2}(1+Mt)^{-1}_{i_3i_3}\nonumber \\
&\cdot &\mathrm{pf}(1+Bt))|_{t^{N/2-2}},
\end{eqnarray}
\begin{eqnarray}
s3_{i_1i_2i_3} &=& ((1+Mt)^{-1}_{i_1i_1}(1+Mt)^{-1}_{i_2i_2}(1+Mt)^{-1}_{i_3i_3}\nonumber \\
&\cdot &\mathrm{pf}(1+Bt))|_{t^{N/2-1}}.
\end{eqnarray}
The above expressions 
show the diagonal elements of 
tensors $q3$, $r3$, and $s3$.
In eq.(\ref{first}),
these tensors are used as 
rank-6 tensors with diagonal entries.
The tensor product with the cross term
is defined similar to that in ref.\cite{uemura201901}.
For example, the first line
\begin{equation}
(Q_rM\times Q_rM\times Q_rM\, q3\, Q_l\times Q_l\times Q_l)_{jl_1l_2ik_2k_1}
\end{equation}
is equal to
\begin{eqnarray}
& &((B(1+Bt)^{-1})_{ji}(B(1+Bt)^{-1})_{l_1k_2}(B(1+Bt)^{-1})_{l_2k_1}\nonumber \\ 
& &\cdot \,\mathrm{pf}(1+Bt))|_{t^{N/2-3}}.
\end{eqnarray}
These values for both the energy
and the first-order derivatives
can be obtained by operation 
with $O(K^2M^5)$ steps,
where $K$ is the number of terms 
of the AGP states and $M$ is the 
number of orbitals or the system size.\par
In the foregoing studies, we refer to 
the variation with the linear
combination of AGP states
as the extended symmetric tensor decomposition (ESTD).
However, we prefer to call it
AGP-CI hereafter, for the sake of clarity.\par
In the real calculation of AGP-CI,
we first consider the variation independently for each
geminal until the 
energy level of the Hartree-Fock is attained.
After all of the geminals are at the Hartree-Fock level, we combine
each geminal state and perform the variation for the total state.\par
In the expression of the first derivative, 
there appear rank-6 tensors, but 
they behave essentially as rank-3 tensors in the real calculations.
The bottleneck of the total variational 
calculation is the product of 
the rank-4 tensor with rank-2 tensors.
We obtained a format to perform this $O(M^5)$ operation for only four times for both the total energy and the first-order derivative using the developed method.
In the future, we intend to further reduce the computational cost for this $O(M^5)$ operation.\par
For the AGP-CI calculation of the 
Hubbard model, we can use the 
tensor decomposition for the 
two-electron energy term given by
\begin{equation}
W_{i_1j_1i_2j_2}=\sum _{i=1}^{M}w^{(1), i}_{i_1}w^{(2), i}_{j_1}w^{(3), i}_{i_2}w^{(4), i}_{j_2},
\end{equation}
where $w^{(1)}$, $w^{(2)}$, $w^{(3)}$, and $w^{(4)}$ are rank-one tensors for the Hubbard repulsion term.
With this decomposition, we found that 
the energy variation could be performed with scaling $O(M^3)$ for the energy
and $O(M^4)$ for the first derivative 
for both the kinetic and 
repulsion energy terms.

\section{Result}

We changed the variational parameters of the geminals in the AGP-CI from the real skew-symmetric matrices to complex skew-symmetric matrices.
As a result, the energy obtained for 
the system of the water molecule 
with STO-3G basis set that we used in 
ref.\cite{uemura201901} came 
closer to the exact value.
Table \ref{tb:h2o_sto3g} shows the result with complex parameters.

\begin{table}[tbh]
\begin{center}%
\begin{tabular}
[c]{|l|l|}\hline
Method & Total energy\\\hline
AGP-CI, real parameters, $K=16$ & -75.012415900\\
AGP-CI, complex parameters, $K=16$ & -75.012425253\\\hline
Exact & -75.012425818\\
\hline
\end{tabular}
\end{center}
\caption{Total energy (in units of hartrees) of H$_{2}$O with 
STO-3G basis set obtained by AGP-CI. For
comparison, the full-CI calculation was taken from ref.\cite{uemura201901}. }%
\label{tb:h2o_sto3g}%
\end{table}

Here, the result with real parameters and the full-CI result was taken from ref.\cite{uemura201901}.
We observed that after changing the parameters to be complex, the resulting energy is closer to the exact energy.
For this case, the resultant energy is around $5.7 \times 10^{-7}$ hartree. 
If we desire, we could just change the threshold value in the variational procedure and could reduce the resultant energy almost arbitrarily.\par
Next, we show the result for the water molecule with the Dunning double-zeta basis set, which is defined in ref.\cite{uemura201901}.
We optimized our numerical code to obtain the energy.
After that, we enlarged the number of states of AGP-CI by almost threefold.
\begin{table}[tbh]
\begin{center}%
\begin{tabular}
[c]{|l|l|}\hline
Method & Total energy\\\hline
Hartree-Fock & -76.00983760\\\hline
AGP-CI, $K=4$ & -76.13814833\\
AGP-CI, $K=60$ & -76.15509884\\
AGP-CI, $K=200$ & -76.15771517\\
\hline
CISD (ref.\cite{saxe}) & -76.150015\\
CCSD (ref.\cite{purvis}) & -76.156078\\
Exact (ref.\cite{saxe}) & -76.157866\\\hline
\end{tabular}
\end{center}
\caption{Total energy (in units of hartrees) of H$_{2}$O 
with the DZ basis set obtained by AGP-CI. For
comparison, the full-CI result was taken from ref.\cite{saxe}.
The CISD result in ref.\cite{saxe} and the CCSD
result in ref.\cite{purvis} are also included.}%
\label{tb:h2o_dzbasis}%
\end{table}
Table \ref{tb:h2o_dzbasis} shows the result of AGP-CI with $K=200$.
All the other values are taken from the references mentioned above.
The configuration interaction singles and doubles (CISD) and the coupled cluster singles and doubles (CCSD) results are also included from the references.
For $K=200$, the resultant energy is around $1.5\times 10^{-4}$ hartree and significantly reduced from the former result.
We understand that this result 
satisfies the requirement of the chemical accuracy.
The size of the Hilbert space of this system is approximately $1.3\times 10^7$; our compact AGP states with $200$ terms well capture the total wavefunction of this size.
\begin{figure}
[ptb]
\begin{center}
\includegraphics[bb=0 0 400 300 height=2.0in, width=3.0in]
{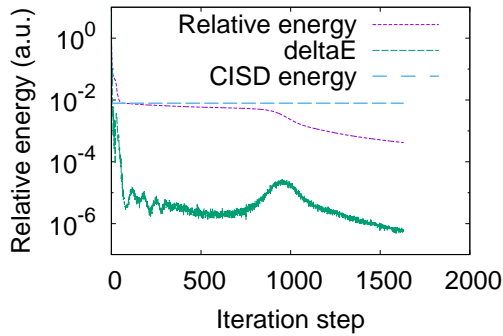}
\caption{The relative energy and energy changes (deltaE) on the variation of 
the water molecule with the DZ basis set. The number of AGPs ($K$) is set to 200.}
\label{figm28h2o}
\end{center}
\end{figure}
Fig.\ref{figm28h2o} shows the behavior of the AGP-CI energy with respect to the exact energy for the iteration steps.
The energy change for one iteration step is also shown as deltaE.
This graph shows the first half of the total iteration steps for this system.
The energy level of CISD is also shown
as a horizontal line. 
As can be seen from the graph, the relative energy first drops sharply and is then trapped near the CISD energy.
After a long interval, the energy starts to drop again and approaches the exact value.
We have not yet clarified the reason why the AGP-CI energy is trapped near the CISD energy.
The reason could be that there is a large concentration of the density of states of the AGPs near the CISD energy level.
We could possibly imagine that if we can construct the CISD state before AGP-CI and prepare the initial AGP-CI wavefunction as imitating the CISD wavefunction then we could improve the quality of the total variation.
Moreover, we could observe the following character of AGP-CI.
For this system, the number of configurations for CISD is approximately the square of the number of orbitals.
Then, since the AGPs include the Slater determinants, we could obtain fairly good AGP-CI results after setting the number of states larger than the square of the number of orbitals.
This would mean that the AGP-CI result is likely to behave polynomially on the computational cost.
As AGP states are continuous functions rather than discrete Slater determinants, we expect good energy improvement compared with CISD.

\par
Thirdly, we introduce the results of AGP-CI for the one-dimensional half-filled Hubbard model with 14 sites.
The parameter $U/t$ is set to $1.0$.
\begin{table}[tbh]
\begin{center}%
\begin{tabular}
[c]{|l|l|}\hline
Method & Total energy\\\hline
Hartree-Fock & -14.47583682\\\hline
AGP-CI, $K=1$ & -14.52254127\\
AGP-CI, $K=4$ & -14.63322447\\
AGP-CI, $K=16$ & -14.70368298\\
AGP-CI, $K=100$ & -14.70437808\\
AGP-CI, $K=500$ & -14.71456970\\\hline
Exact (ref.\cite{booth}) & -14.7147075\\\hline
\end{tabular}
\end{center}
\caption{Total energy of the 14-site Hubbard model with 
U/t=1.0 obtained by AGP-CI. For
comparison, the full-CI result of ref.\cite{booth} is also shown.}%
\label{tb:hubbard_s14}%
\end{table}
Table \ref{tb:hubbard_s14} shows our result of AGP-CI for this system.
The result with $K=1$ is slightly below the Hartree-Fock energy.
When we increase the number of states up to $500$, we observe a resultant energy of $1.3\times 10^{-4}$.
We understand that this result is in good agreement with the right ground state.
This size of the number of states in AGP-CI was possible to achieve after introducing the Hamiltonian decomposition for the Hubbard repulsion term. 
\begin{figure}
[ptb]
\begin{center}
\includegraphics[bb=0 0 400 300 height=2.0in, width=3.0in]
{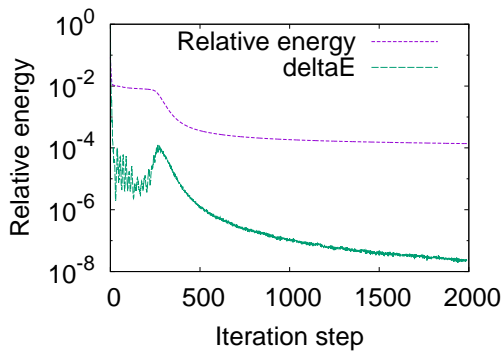}
\caption{The relative energy and energy changes on the variation of 
the one-dimensional 14-site Hubbard model with $U/t = 1.0$. The number of AGPs ($K$) is set to 500.}
\label{figm28u1s14}
\end{center}
\end{figure}
Fig.\ref{figm28u1s14} shows the behavior of the energy of this system on the variation process.
When the variation starts, the energy is trapped on some local energy level.
Then, after some interval, the energy rapidly decreases and becomes closer to the exact value.
Then, the variation is almost converged.
For this system the size of the total Hilbert space is about $4.0 \times 10^{7}$. The ground-state wavefunction of this large Hilbert space is well captured by $500$ AGP states.\par
Finally, we introduce the result for the $4 \times 4$ two-dimensional half-filled Hubbard model.
The parameter $U/t$ is set to $10.0$.
\begin{table}[tbh]
\begin{center}%
\begin{tabular}
[c]{|l|l|}\hline
Method & Total energy\\\hline
Hartree-Fock & -6.06641304\\\hline
AGP-CI, $K=1$ & -6.08549342\\
AGP-CI, $K=4$ & -6.46641223\\
AGP-CI, $K=16$ & -6.68276779\\
AGP-CI, $K=100$ & -6.91808110\\
AGP-CI, $K=400$ & -7.00753607\\\hline
Exact (ref.\cite{imada}) & -7.13239\\\hline
\end{tabular}
\end{center}
\caption{Total energy of the $4 \times 4$ Hubbard model with 
U/t=10.0 obtained by AGP-CI. For
comparison, the full-CI result of ref.\cite{imada} is also shown.}%
\label{tb:hubbard_2d}%
\end{table}
In Table \ref{tb:hubbard_2d}, we show the AGP-CI result for this system.
The energy gradually improves when 
the number of states increases.
When $K=400$, the resultant energy is around $0.13$.
This two-dimensional Hubbard model with a relatively large repulsion energy seems to be strongly correlated, and a large number of states are required for a better description of the ground state.
In this system, the dimension of the total Hilbert space is approximately $6.0 \times 10^8$, which is larger than the former systems.
\begin{figure}
[ptb]
\begin{center}
\includegraphics[bb=0 0 400 300 height=2.0in, width=3.0in]
{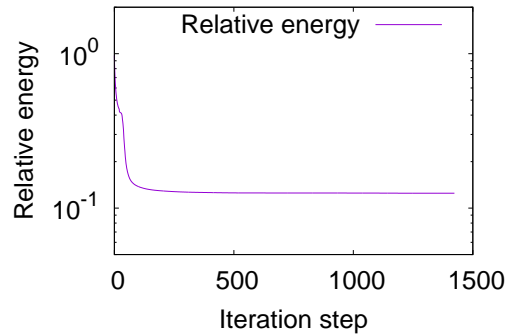}
\caption{The relative energy on the variation of the $4 \times 4$ Hubbard model with $U/t = 10.0$. The number of AGPs ($K$) is set to 400.}
\label{figm32u10s16}
\end{center}
\end{figure}
Fig.\ref{figm32u10s16} shows the behavior of the relative energy on the variation process.
As we can see from the figure, the energy sharply drops down on the early stage of the variation.
Then, after the energy attains a certain value, the variation converges.
This system of the two-dimensional Hubbard model has a highly symmetric structure; therefore, if we could include this symmetry of the Hubbard model, the AGP-CI variation would dramatically improve.\par
\begin{figure}
[ptb]
\begin{center}
\includegraphics[bb=0 0 400 250 height=2.0in, width=3.0in]
{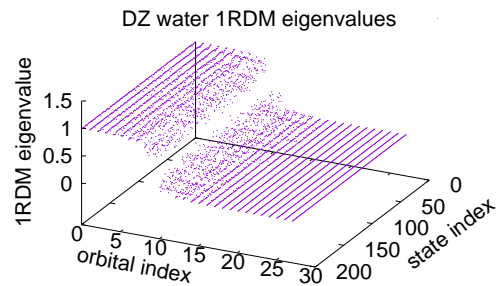}
\caption{1RDM eigenvalues of 
the water molecule with the DZ basis set. 
Eigenvalues are taken for each AGP state in AGP-CI.
The number of AGPs ($K$) is set to 200, and the number of electrons is set to 10.}
\label{figm28h2o_eigen}
\end{center}
\end{figure}
\begin{figure}
[ptb]
\begin{center}
\includegraphics[bb=0 0 400 300 height=2.2in, width=3.0in]
{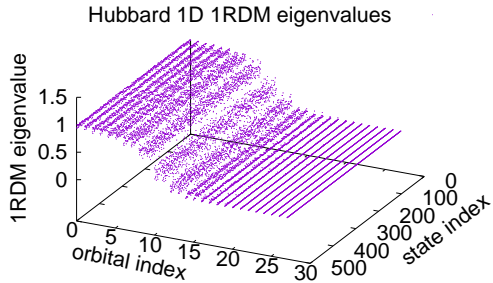}
\caption{1RDM eigenvalues of 
the one-dimensional 14-site Hubbard model with $U/t = 1.0$. The number of AGPs ($K$) is set to 500, and
the number of electrons is set to 14.}
\label{figm28u1s14_eigen}
\end{center}
\end{figure}
\begin{figure}
[ptb]
\begin{center}
\includegraphics[bb=0 0 400 300 height=2.2in, width=3.0in]
{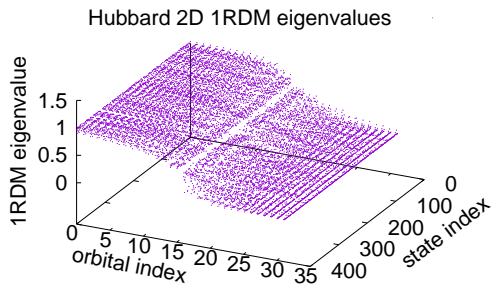}
\caption{1RDM eigenvalues of the $4 \times 4$ Hubbard model with $U/t = 10.0$. The number of AGPs ($K$) is set to 400, and
the number of electrons is set to 16.}
\label{figm32u10s16_eigen}
\end{center}
\end{figure}
We further analyzed the distribution of the 
eigenvalues of the first-order reduced density matrix (1RDM)
for each AGP state in AGP-CI.
Figs. \ref{figm28h2o_eigen}, \ref{figm28u1s14_eigen}, and \ref{figm32u10s16_eigen} respectively show the eigenvalue distribution for the DZ water, 
the one-dimensional Hubbard model, and the two-dimensional Hubbard model 
at the end of the variation.
The horizontal plane shows the index of the AGP state
and the index of the eigenvalue.
The vertical axis shows the magnitude of the eigenvalue.
The eigenvalue of the 1RDM corresponds to the occupation number of the natural orbital of the AGP state.
These distribution of the eigenvalue does not appear on the case of Slater determinants.
As we can see from these figures, the distribution of 
the eigenvalue has a similar shape between different
geminals in each system.
This is a result of the condition of the AGP state that each AGP state in AGP-CI
should have energy close to the ground-state energy.
The distribution of DZ water is most narrow and
that of the two-dimensional Hubbard model is most broad 
in these three cases.
The broadness of these distributions might be
reflecting the strength of the electronic 
correlation of each system.
Therefore, we can assume that the requirement for a larger number of states for the two-dimensional Hubbard model
is a result of the stronger electronic correlation.
Since these distributions are similar for each AGP state, we can assume that the required geminals in AGP-CI
are distributing across a restricted area of the geminal space
characterized by each system.
It would benefit the whole calculation if we 
successfully characterize the nature of the geminals 
mentioned above.
This could be done with a well-organized control of the eigenvalues of each geminal matrix.

\section{Conclusion}

We tested the AGP-CI formalism
on a double-zeta water molecule and Hubbard models with an extended number of sites.
The results for the water molecule and 14-site Hubbard model were very close to the exact result.
From this, we conclude that the required number of states for AGP-CI increases only moderately with the problem size.
The calculation on the Hubbard model was based on the tensor decomposition for the two-electron part of the Hamiltonian.
We could also use this tensor decomposition for molecular systems
and could possibly reduce the computational scaling by one order.\cite{udo}
There are non-trivial relationships of the AGP-CI energy with the CISD energy.
If we could use the CISD wavefunctions before the AGP-CI calculation, we would expect that the total variation process would become shorter.
In the future, we will further study the dependence of the required number of AGP states on the size of the problem.
If we could utilize tensor decomposition for the Hamiltonian, we could investigate the behavior of AGP-CI for larger systems. 
With an analysis of the 
nature of the geminal states appearing in AGP-CI, we can improve the whole variation process.
We only used dozens of CPUs 
for the above calculations; however, if this is extended to larger computing resources, we can perform the AGP-CI calculation with more extended systems.

\section*{Acknowledgment}

We acknowledge the Strategic Programs for Innovative
Research by MEXT and ``Priority Issue on Post-K Computer''
(Development of new fundamental technologies for high-efficiency
energy creation, conversion or storage, and use) for
financial support during our research. Some of the computations
in the present study were performed at the Institute for
Molecular Science, Okazaki, Japan.


\bigskip

\begin{thebibliography}{99}                                                                                               

\bibitem{shavitt}I. Shavitt, Mol. Phys. 94, 3 (1998).

\bibitem{saxe}P. Saxe, H. F. Schaefer, and N. C. Handy, Chem. Phys. Lett. 79, 202
(1981).

\bibitem{purvis}G. D. Purvis and R. J. Bartlett, J. Chem. Phys. 76,
1910(1982).

\bibitem{knowles}H. Werner and P. J. Knowles, J. Chem. Phys. 89, 5803 (1988).

\bibitem{Adamowicz}N. Oliphant and L. Adamowicz, J. Chem. Phys. 94, 1229 (1991).

\bibitem{evangelista}F. A. Evangelista, J. Chem. Phys. 149, 030901 (2018).

\bibitem{hohenberg}P. Hohenberg and W. Kohn, Phys. Rev. 136, B864 (1964).

\bibitem{kohn}W. Kohn and L. J. Sham, Phys. Rev. 140, A1133 (1965).

\bibitem{levy}M. Levy, Phys. Rev. A 26, 1200 (1982).

\bibitem {foulkes}W. M. C. Foulkes, L. Mitas, R. J. Needs, and G. Rajagopal,
Rev. Mod. Phys. 73, 33 (2001).

\bibitem {bajdich2006}M. Bajdich, L. Mitas, G. Drobny, L. K. Wagner, and K. E.
Schmidt, Phys. Rev. Lett. 96, 130201 (2006).

\bibitem {bajdich2008}M. Bajdich, L. Mitas, L. K. Wagner, and K. E. Schmidt,
Phys. Rev. B 77, 115112 (2008).

\bibitem {eric}E. Neuscamman, Phys. Rev. Lett. 109, 203001 (2012).

\bibitem{eric2}E. Neuscamman, J. Chem. Phys. 139, 194105 (2013).

\bibitem {sorella}A. Zen, E. Coccia, Y. Luo, S. Sorella, and L. Guidoni, J.
Chem. Theory Comput. 10 (3), 1048 (2014).

\bibitem {sorella2}Y. Luo, A. Zen, and S. Sorella, J. Chem. Phys. 141, 194112 (2014).

\bibitem {sorella3}A. Zen, Y. Luo, G. Mazzola, L. Guidoni, and S. Sorella, J. Chem. Phys. 142,
144111 (2015).

\bibitem{booth}N. S. Blunt, S. D. Smart, G. H. Booth, and A. Alavi,
J. Chem. Phys. 143, 134117 (2015).

\bibitem{haoshi}H. Shi and S. Zhang, Phys. Rev. B 95, 045144 (2017).

\bibitem{imada}D. Tahara and M. Imada, J. Phys. Soc. Jpn. 77, 114701 (2008).

\bibitem{fukutome}A. Ikawa, S. Yamamoto, and H. Fukutome, J. Phys. Soc. Jpn. 62, 1653 (1993).

\bibitem{tomita1}N. Tomita, Phys. Rev. B 69, 045110 (2004).

\bibitem{tomita2}N. Tomita, Phys. Rev. B 79, 075113 (2009).

\bibitem{tomita3}S. Watanabe, M. Katoh, and N. Tomita, J. Phys. Soc. Jpn. 82, 044705 (2013).

\bibitem{hoyos1}R. Rodriguez-Guzman, C. A. Jimenez-Hoyos, R. Schutski, and G. E. Scuseria, Phys. Rev. B 87, 235129 (2013).

\bibitem{hoyos2}C. A. Jimenez-Hoyos, R. Rodriguez-Guzman, and G. E. Scuseria, J. Chem. Phys. 139, 204102 (2013).

\bibitem{hoyos3}R. Rodriguez-Guzman, C. A. Jimenez-Hoyos, and G. E. Scuseria, Phys. Rev. B 90, 195110 (2014).

\bibitem {coleman1}A. J. Coleman, Rev. Mod. Phys. 35, 668 (1963).

\bibitem {coleman2}A. J. Coleman, J. Math. Phys. 6, 1425 (1965).

\bibitem {Ortiz1981}J. V. Ortiz, B. Weiner, and Y. Ohn, Int. J. Quantum Chem. 20,
113 (1981).

\bibitem {weiner}B. Weiner and J. V. Ortiz, J. Chem. Phys. 117, 5135 (2002).

\bibitem{giorgini}S. Giorgini, L. P. Pitaevskii, and S. Stringari, Rev. Mod. Phys. 80, 1215-1274 (2008).

\bibitem {Straroverov2002}V. N. Starovecrov and G. E. Scuseria, J. Chem. Phys.
117, 11107 (2002).

\bibitem {Scuseria2011}G. E. Scuseria, C. A. Jimenez-Hoyos, T. M. Henderson,
K. Samanta, and J. K. Ellis, J. Chem. Phys. 135, 124108 (2011).

\bibitem{uemura2015}W. Uemura, S. Kasamatsu, and O. Sugino, Phys. Rev. A 91, 062504 (2015).

\bibitem{uemura201901}W. Uemura and 
T. Nakajima, Phys. Rev. A 99, 012519 (2019).

\bibitem{mizusaki}T. Mizusaki, M. Oi, and N. Shimizu, Phys. Lett. B779, 237(2018). 

\bibitem{udo}U. Benedikt, A. A. Auer, M. Espig, and W. Hackbusch, J. Chem. Phys. 134, 054118 (2011).





\end{thebibliography}
\end{document}